\documentclass[ALICE,manyauthors]{cernphprep}
\usepackage[comma,square,numbers,sort&compress]{natbib}
\usepackage{hyperref}
\usepackage{lineno}
\usepackage{xspace}
\usepackage{graphicx}
\usepackage{xcolor}
\usepackage{float}

\usepackage[T1]{fontenc}
\usepackage{orcidlink}

\begin{document}

%

\newcommand{\pp}           {pp\xspace}
\newcommand{\ppbar}        {\mbox{$\mathrm {p\overline{p}}$}\xspace}
\newcommand{\XeXe}         {\mbox{Xe--Xe}\xspace}
\newcommand{\PbPb}         {\mbox{Pb--Pb}\xspace}
\newcommand{\pA}           {\mbox{pA}\xspace}
\newcommand{\pPb}          {\mbox{p--Pb}\xspace}
\newcommand{\AuAu}         {\mbox{Au--Au}\xspace}
\newcommand{\dAu}          {\mbox{d--Au}\xspace}

\newcommand{\s}            {\ensuremath{\sqrt{s}}\xspace}
\newcommand{\snn}          {\ensuremath{\sqrt{s_{\mathrm{NN}}}}\xspace}
\newcommand{\pt}           {\ensuremath{p_{\rm T}}\xspace}
\newcommand{\meanpt}       {$\langle p_{\mathrm{T}}\rangle$\xspace}
\newcommand{\ycms}         {\ensuremath{y_{\rm CMS}}\xspace}
\newcommand{\ylab}         {\ensuremath{y_{\rm lab}}\xspace}
\newcommand{\etarange}[1]  {\mbox{$\left | \eta \right |~<~#1$}}
\newcommand{\yrange}[1]    {\mbox{$\left | y \right |~<~#1$}}
\newcommand{\dndy}         {\ensuremath{\mathrm{d}N_\mathrm{ch}/\mathrm{d}y}\xspace}
\newcommand{\dndeta}       {\ensuremath{\mathrm{d}N_\mathrm{ch}/\mathrm{d}\eta}\xspace}
\newcommand{\avdndeta}     {\ensuremath{\langle\dndeta\rangle}\xspace}
\newcommand{\dNdy}         {\ensuremath{\mathrm{d}N_\mathrm{ch}/\mathrm{d}y}\xspace}
\newcommand{\Npart}        {\ensuremath{N_\mathrm{part}}\xspace}
\newcommand{\Ncoll}        {\ensuremath{N_\mathrm{coll}}\xspace}
\newcommand{\dEdx}         {\ensuremath{\textrm{d}E/\textrm{d}x}\xspace}
\newcommand{\RpPb}         {\ensuremath{R_{\rm pPb}}\xspace}

\newcommand{\nineH}        {$\sqrt{s}~=~0.9$~Te\kern-.1emV\xspace}
\newcommand{\seven}        {$\sqrt{s}~=~7$~Te\kern-.1emV\xspace}
\newcommand{\twoH}         {$\sqrt{s}~=~0.2$~Te\kern-.1emV\xspace}
\newcommand{\twosevensix}  {$\sqrt{s}~=~2.76$~Te\kern-.1emV\xspace}
\newcommand{\five}         {$\sqrt{s}~=~5.02$~Te\kern-.1emV\xspace}
\newcommand{\twosevensixnn}{$\sqrt{s_{\mathrm{NN}}}~=~2.76$~Te\kern-.1emV\xspace}
\newcommand{\fivenn}       {$\sqrt{s_{\mathrm{NN}}}~=~5.02$~Te\kern-.1emV\xspace}
\newcommand{\LT}           {L{\'e}vy-Tsallis\xspace}
\newcommand{\GeVc}         {Ge\kern-.1emV/$c$\xspace}
\newcommand{\MeVc}         {Me\kern-.1emV/$c$\xspace}
\newcommand{\TeV}          {Te\kern-.1emV\xspace}
\newcommand{\GeV}          {Ge\kern-.1emV\xspace}
\newcommand{\MeV}          {Me\kern-.1emV\xspace}
\newcommand{\GeVmass}      {Ge\kern-.2emV/$c^2$\xspace}
\newcommand{\MeVmass}      {Me\kern-.2emV/$c^2$\xspace}
\newcommand{\lumi}         {\ensuremath{\mathcal{L}}\xspace}

\newcommand{\ITS}          {\rm{ITS}\xspace}
\newcommand{\TOF}          {\rm{TOF}\xspace}
\newcommand{\ZDC}          {\rm{ZDC}\xspace}
\newcommand{\ZDCs}         {\rm{ZDCs}\xspace}
\newcommand{\ZNA}          {\rm{ZNA}\xspace}
\newcommand{\ZNC}          {\rm{ZNC}\xspace}
\newcommand{\SPD}          {\rm{SPD}\xspace}
\newcommand{\SDD}          {\rm{SDD}\xspace}
\newcommand{\SSD}          {\rm{SSD}\xspace}
\newcommand{\TPC}          {\rm{TPC}\xspace}
\newcommand{\TRD}          {\rm{TRD}\xspace}
\newcommand{\VZERO}        {\rm{V0}\xspace}
\newcommand{\VZEROA}       {\rm{V0A}\xspace}
\newcommand{\VZEROC}       {\rm{V0C}\xspace}
\newcommand{\Vdecay} 	   {\ensuremath{V^{0}}\xspace}

\newcommand{\ee}           {\ensuremath{e^{+}e^{-}}} 
\newcommand{\pip}          {\ensuremath{\pi^{+}}\xspace}
\newcommand{\pim}          {\ensuremath{\pi^{-}}\xspace}
\newcommand{\kap}          {\ensuremath{\rm{K}^{+}}\xspace}
\newcommand{\kam}          {\ensuremath{\rm{K}^{-}}\xspace}
\newcommand{\pbar}         {\ensuremath{\rm\overline{p}}\xspace}
\newcommand{\kzero}        {\ensuremath{{\rm K}^{0}_{\rm{S}}}\xspace}
\newcommand{\lmb}          {\ensuremath{\Lambda}\xspace}
\newcommand{\almb}         {\ensuremath{\overline{\Lambda}}\xspace}
\newcommand{\Om}           {\ensuremath{\Omega^-}\xspace}
\newcommand{\Mo}           {\ensuremath{\overline{\Omega}^+}\xspace}
\newcommand{\X}            {\ensuremath{\Xi^-}\xspace}
\newcommand{\Ix}           {\ensuremath{\overline{\Xi}^+}\xspace}
\newcommand{\Xis}          {\ensuremath{\Xi^{\pm}}\xspace}
\newcommand{\Oms}          {\ensuremath{\Omega^{\pm}}\xspace}
\newcommand{\degree}       {\ensuremath{^{\rm o}}\xspace}

\newcommand{\gab}          {\ensuremath{\langle \cos(\varphi_\alpha + \varphi_\beta - 2\psid) \rangle}\xspace}

\newcommand{\dab}          {\ensuremath{\langle \cos(\varphi_\alpha - \varphi_\beta) \rangle}\xspace}

\newcommand{\gc}           {\ensuremath{\gamma_{\alpha\beta}}\xspace}

\newcommand{\dc}         {\ensuremath{\delta_{\alpha\beta}}\xspace}
\newcommand{\psid}         {\ensuremath{\Psi_{\rm 2}}\xspace}

\newcommand{\fcme}         {\ensuremath{f_{\rm CME}}\xspace}
\newcommand{\fcmexe}         {\ensuremath{f_{\rm CME}^{\rm Xe-Xe}}\xspace}
\newcommand{\fcmepb}         {\ensuremath{f_{\rm CME}^{\rm Pb-Pb}}\xspace}

\begin{titlepage}
\PHyear{2026}       
\PHnumber{035}      
\PHdate{13 Feb}  

\title{Limits on the chiral magnetic effect from the event shape engineering
and participant-spectator correlation techniques in Pb--Pb collisions at $\bf{\snn = 5.02}$~TeV}
\ShortTitle{Chiral magnetic effect search at the LHC}   

\Collaboration{ALICE Collaboration\thanks{See Appendix~\ref{app:collab} for the list of collaboration members}}
\ShortAuthor{ALICE Collaboration} 

\begin{abstract}
The latest experimental studies related to the search for the Chiral Magnetic Effect (CME) in 
Pb--Pb collisions at $\snn=5.02$~TeV~recorded with the ALICE detector at the Large Hadron Collider (LHC) are 
presented. Charge-dependent two-particle correlations relative to the reaction plane are measured for charged particles in the pseudorapidity 
range $|\eta| < 0.8$ and the transverse-momentum range 
$0.2< p_{\mathrm{T}} < 5$~GeV/$c$. Two approaches have been employed: in the 
first method, the contribution of the background to the measurement is varied using the event shape engineering (ESE), while the second relies on changing the contribution of the potential CME signal by measuring azimuthal correlations relative to the participant plane, 
where the background contributions are maximized, and spectator plane, where the 
CME signal contribution is maximized. 
Both methods yield 
results consistent with the absence of a CME signal within the measurement uncertainties. The result obtained from correlations relative to different symmetry planes, a technique applied for the first time at LHC energies, gives the possibility to test independently and confirm the upper limits from previous measurements, while the new limit from the ESE analysis offers improved constraint relative to previous attempts.  
\end{abstract}
\end{titlepage}

\setcounter{page}{2} 


\section{Introduction}
\label{Sec:Intro}

Ultra-relativistic heavy-ion collisions create the necessary conditions for the formation of a quark--gluon plasma (QGP), a state of matter in which quarks and gluons are deconfined 
and chiral symmetry is restored as a result of extreme temperatures and energy 
densities~\cite{Shuryak:1978ij,Karsch:2003jg,Bazavov:2009zn}. Theory expects that in a 
chirally restored medium such as the QGP~\cite{HotQCD:2018pds}, local parity violation could occur due to 
transitions of topologically non-trivial configurations of the gluonic field, coined as 
instantons and sphalerons, leading to a local imbalance of quark 
chirality~\cite{Morley:1983wr,Kharzeev:1998kz,Kharzeev:1999cz,Kharzeev:2007tn,Kharzeev:2007jp,Kharzeev:2015kna}. 
In the initial-state of non-central collisions, spectator protons, i.e., those not participating directly in the interaction, generate extremely strong magnetic fields with magnitudes reaching up to $10^{15}$~T~\cite{Skokov:2009qp,Bzdak:2011yy,Deng:2012pc}. While the existence of these magnetic fields is theoretically well established, there is currently no direct experimental evidence for their effect on the motion of final-state particles since the field's magnitude is expected to decay rapidly~\cite{Toneev:2011aa,Shi:2017cpu}. The direction of the magnetic field is, on average, perpendicular to the reaction
plane, which is defined by the impact parameter vector and the beam axis. In the presence
of the initial-state magnetic field, the chirality imbalance could lead to the development
of an electric current perpendicular to the reaction plane and parallel to the magnetic field. 
This phenomenon is known as the Chiral Magnetic Effect (CME)~\cite{Fukushima:2008xe, Kharzeev:2007jp} and is closely related to other anomalous phenomena such as the Chiral Separation Effect~(CSE)~\cite{Son:2009tf}, the Chiral Magnetic
Wave (CMW)~\cite{Kharzeev:2010gd,Burnier:2011bf}, and the Chiral Vortical Effect (CVE)~\cite{Kharzeev:2010gr}.  
The CME manifests itself as a charge separation in the azimuthal distribution of produced 
particles relative to the reaction plane. The dominant component in the Fourier expansion 
of this distribution arises from anisotropic flow, a phenomenon originating from the initial-state asymmetry of the collision in coordinate space that is transformed into an anisotropy 
in momentum space via interactions between partons and at later stages between the produced 
particles. Anisotropic flow is usually characterized by the Fourier coefficients $v_n$ in the 
expansion of the azimuthal distribution of produced particles~\cite{Voloshin:1994mz,Poskanzer:1998yz}, 
with the first- ($v_1$) and second-order ($v_2$) coefficients corresponding to the directed 
and elliptic (the dominant harmonic) flow, respectively. In order to probe and quantify these 
local parity violation effects, one may introduce an additional sine component to this 
distribution, which takes the form
\begin{equation}
\label{Eq:dNdPhiFlowAndCME}
\frac{dN}{d\varphi_i}\sim 1+2\sum_{n=1}^\infty [a_{n,i}\sin{[n(\varphi_i-\Psi_{\text{RP}})]}+ v_{n,i}\cos{[n(\varphi_i-\Psi_{\text{RP}})]},
\end{equation}
where $\varphi_i$ represents the azimuthal angle of particle $i$ and $\Psi_{\mathrm{RP}}$ 
denotes the reaction plane angle. Due to fluctuations in the positions of nucleons within the colliding nuclei, the actual geometry of the overlapping region varies from one event to another~\cite{Bhalerao:2006tp,PHOBOS:2007vdf,Alver:2010gr,PHOBOS:2005gex}. This gives rise to additional symmetry planes, such as the participant plane ($\Psi_{\mathrm{PP}}$), defined by the anisotropic distribution of participating nucleons and the spectator plane ($\Psi_{\mathrm{SP}}$), determined by the deflected spectator nucleons. These planes are correlated to the reaction plane but fluctuate around it event-by-event. In Eq.~\ref{Eq:dNdPhiFlowAndCME}, the first-order coefficient $a_{1,i}$ quantifies the magnitude of the local parity violation effect. Due to the spontaneous nature of the topological transitions, the effect averages to zero over many collision events~\cite{Voloshin:2004vk}. This makes direct observation 
of $a_{1,i}$ challenging. Experimental techniques based on the study of 
charge-dependent azimuthal correlations are consequently developed to detect possible signatures 
of the CME~\cite{Voloshin:2004vk}. These techniques are based on a charge-dependent azimuthal 
correlator 
\begin{equation}
    \gamma_{\alpha\beta}=\langle \cos{(\varphi_{\alpha}+\varphi_{\beta}-2\Psi_{\mathrm{RP}})}\rangle=\langle \cos(\Delta\varphi_\alpha)\cos(\Delta\varphi_\beta)\rangle-\langle \sin(\Delta\varphi_\alpha)\sin(\Delta\varphi_\beta)\rangle,
\end{equation}
where the angular brackets indicate an average over all events, and $\alpha$ and $\beta$ denote 
the particle charge that allows to form particle pairs of opposite or same sign within the same event. The terms $\langle\cos{(\Delta\varphi_{\alpha})}\cos{(\Delta\varphi_{\beta})}\rangle$ 
and $\langle\sin{(\Delta\varphi_{\alpha})}\sin{(\Delta\varphi_{\beta})}\rangle$, with $\Delta\varphi_{\alpha,\beta} = \varphi_{\alpha,\beta} - \Psi_{\mathrm{RP}}$, quantify the 
correlations with respect to the in- and out-of-plane directions, respectively. The CME 
sensitive observable is constructed by taking the difference of opposite-sign pairs $\gamma_{+-}$ (OS) and same-sign pairs $(\gamma_{++} + \gamma_{--})/2$ (SS), and is denoted as $\Delta\gamma$. This 
reduces contributions from background associated with charge-independent sources. 

The first results relying on this approach were published by the STAR Collaboration based on the 
analysis of Au--Au collisions at $\sqrt{s_{\mathrm{NN}}} = 0.2$~TeV~\cite{Abelev:2009ac,Abelev:2009ad}. 
Subsequent analyses~\cite{STAR:2013ksd,STAR:2014uiw,STAR:2013zgu,STAR:2019xzd,STAR:2020gky,STAR:2021pwb} 
confirmed the observation of a result which initially seemed to be consistent with a charge separation relative to the 
reaction plane, in accordance to what is expected due to the CME. Quantitatively 
similar results were also reported at much higher energies at the Large Hadron Collider 
(LHC)~\cite{Abelev:2012pa,Acharya:2020rlz,Khachatryan:2016got,Sirunyan:2017quh}.
It was later recognized that the measured $\Delta\gamma$ correlator contains substantial contribution from non-CME backgrounds~\cite{Voloshin:2004vk,Wang:2009kd,Bzdak:2009fc,Liao:2010nv,Schlichting:2010qia,ALICE:2022ljz,Wu:2022fwz,Christakoglou:2021nhe}. These backgrounds are primarily attributed to local charge conservation (LCC), i.e., the correlated emission of oppositely charged particles from a cluster, occurring within an anisotropically expanding medium, thereby mimicking the CME--like charge separation.

Lacking unambiguous evidence for the CME, the subsequent experimental studies focused 
on two approaches. In the first one, one tries to vary and quantify the background while 
keeping the contribution of the potential CME signal unaltered. Such an approach is embodied in the event 
shape engineering studies that allow to select events with different magnitude of elliptic flow within 
the same centrality~\cite{Schukraft:2012ah}. This gives the possibility to quantify the dependence of 
$\Delta \gamma$ on one of the main background components, i.e., the elliptic flow. In the second approach, one modifies 
the signal contribution to $\Delta\gamma$ and looks for relative changes in the measurement. One example is experiments with isobar collisions at RHIC energies, which allowed to extract
upper limits on the CME contribution to $\Delta \gamma$~\cite{STAR:2021mii, STAR:2023gzg}. 
Another example are measurements of the charge-dependent
correlations relative to different symmetry planes~\cite{Xu:2017qfs,Voloshin:2018qsm}. This can be done 
by calculating the $\gamma$ correlator relative to either $\Psi_{\mathrm{PP}}$ or $\Psi_{\mathrm{SP}}$. The measurement relative to $\Psi_{\mathrm{SP}}$ will potentially have a bigger 
contribution from the CME signal since the charge separation takes place along the magnetic field 
predominantly created by spectator protons. At the same time, since the major background contribution 
comes from LCC convoluted with elliptic flow~\cite{Voloshin:2004vk}, which is a maximum in participant geometry, its 
contribution will be the largest when the azimuthal angles of the two particles are measured relative 
to $\Psi_{\mathrm{PP}}$. Both approaches have been investigated so far and did not yield any significant evidence for the existence of the CME~\cite{Acharya:2017fau, STAR:2025vhs, STAR:2021pwb}.

In this article, the results of the analysis of Pb--Pb collisions at $\snn=5.02$~TeV~recorded 
with ALICE at the LHC and based on the ESE technique and on correlations relative to $\Psi_{\mathrm{PP}}$ 
and $\Psi_{\mathrm{SP}}$ are presented. The article is organised as follows. Section~\ref{Sec:Exp} describes briefly 
the experimental layout and the data sample used, followed by a presentation of the two methods in 
Section~\ref{Sec:Analysis}. The systematic uncertainties of both measurements are discussed in Section~\ref{Sec:Uncertainties}. Section~\ref{Sec:Results} presents the main results together with an overview of the new and previous ALICE findings. The final section contains the conclusion and outlook.

\section{Experimental setup and data sample}
\label{Sec:Exp}
The experimental setup of ALICE consists of a central barrel that contains several detectors and 
a set of forward systems~\cite{Aamodt:2008zz}. Details about the 
various detector sub-systems and their performance can be found in Ref.~\cite{Abelev:2014ffa}. 

Charged-particle tracks are reconstructed with the main tracking detectors, the Inner Tracking System (ITS)~\cite{Aamodt:2010aa} and the Time Projection 
Chamber (TPC)~\cite{Alme:2010ke}, positioned in the 
central barrel inside a solenoid 
magnet which generates up to a 0.5~T field parallel to the beam direction. For the 2018 Pb--Pb data analysed in this work, the ITS, similarly to the first period of ALICE operations,  consisted of six layers of silicon detectors employing 
three different technologies: Silicon Pixel Detectors (SPD) for the two innermost layers, followed 
by two layers of Silicon Drift Detectors (SDD), and two outermost layers with double-sided Silicon 
Strip Detectors (SSD). The SPD layers are also employed for event selection and primary vertex reconstruction. 
The TPC, which surrounds the ITS, is used as the main tracking device as well as for particle 
identification via the measurement of the specific energy loss, d$E$/d$x$. In the forward region, two 
scintillator arrays (V0)~\cite{Abbas:2013taa} are used for triggering, event selection, the 
determination of the collision centrality~\cite{ALICE:2013hur} and participant plane. The V0 consists of two systems, 
the V0C and V0A, covering the pseudorapidity ranges $-3.7 < \eta < -1.7$ and $2.8 < \eta < 5.1$, 
respectively. In addition, two tungsten-quartz neutron Zero Degree Calorimeters (ZDCs), installed 
more than 100~m from the interaction point on each side (ZDC-A and ZDC-C) and positioned at zero degree with respect to the LHC axis, are used for event 
selection and for the reconstruction of the spectator plane~\cite{Arnaldi:1999zz}.

Approximately 235 million minimum-bias Pb--Pb events at $\snn=5.02$~TeV~recorded in 2018 are 
selected and analysed. The minimum-bias trigger is provided by requiring signals in both V0A 
and V0C scintillator arrays. An additional sample of central and semicentral collisions is selected online based on the V0 signal amplitudes, which enhanced the number of events in the 0--10\% and 30--50\% centrality ranges by about seven and three times relative to the minimum-bias sample, respectively. Beam-induced background events (i.e., 
beam–gas interactions) are removed offline utilizing the V0 and ZDC timing information~\cite{Abelev:2014ffa}. Pileup 
of collisions from different bunch crossings in the TPC is rejected by comparing multiplicity 
estimates from the V0 detector to those of tracking detectors in the central barrel, profiting 
from the difference in readout times between the systems. Same-bunch collision pileup is negligible in the Pb--Pb sample. The primary vertex position, determined 
from tracks reconstructed in the ITS and TPC, is required to be within 10~cm from the nominal 
interaction point along the beam direction. Events are classified in centrality intervals using 
the V0 detector; those in the 0--60\% range of the total hadronic cross section are selected for 
the ESE analysis, while those in the 10--50\% range are used for the participant/spectator-plane 
analysis. Residual systematic effects in the estimation of the spectator plane with the ZDC, related to beam optics configuration used during Pb--Pb data taking, were the reason for the difference in the centrality range used for the two approaches.

Primary charged particles reconstructed with the combination of the ITS and the TPC are required 
to be within the pseudorapidity range of $|\eta| < 0.8$, which avoids effects at
the edges of the detector acceptance. The transverse momentum is chosen to be 
within the range of $0.2 < \pt < 5.0$~GeV/$c$ for the ESE analysis and $0.2 < \pt < 3.0$~GeV/$c$ 
for the participant/spectator-plane analysis, respectively. These two $\pt$ intervals have no influence on the final conclusion reported in this article. Results with the ESE method for $0.2 < \pt < 3.0$~GeV/$c$ did not show any significant difference relative to those reported in this article. Each charged-particle track is required 
to have a minimum number of 70 TPC space points out of a maximum of 159, a 
minimum number of crossed TPC pad rows of 70, an expected fraction of TPC space points to reconstruct a
track of 0.8 and a $\chi^2$ per TPC 
space point less than 2.5. In addition, tracks are selected if they had at least 2 hits in the six 
ITS layers, with a $\chi^2$ per ITS hit less than 36. To ensure that secondary particles, either 
from a weak decay of primaries or from the interaction of particles with the detector material, 
are removed, a distance of closest approach (DCA) from the primary vertex smaller than 2~cm in the longitudinal direction is required. An additional DCA selection in the transverse plane was further applied based on a \pt-dependent maximum value of the form $0.0105 + 0.0350~\pt^{-1.1}$~cm.  
These selection criteria lead to an efficiency for primary tracks that increases from about 70\% at $\pt=0.2$~GeV/$c$ to about 80\% at $\pt=1$~GeV/$c$, and a contamination by secondary charged particles that ranges from about 8.5\% at $\pt=0.2$~GeV/$c$ to about 1\% for $\pt>2$~GeV/$c$~\cite{ALICE:2016ccg}.
All results reported below are corrected for detector inefficiencies.

\section{Analysis methods}
\label{Sec:Analysis}

For both analyses, the particles of interest are selected within the acceptance of the \TPC. For the event shape engineering method, the $q$-vectors are calculated using hits within the acceptance of the V0 detectors. The same detectors are used in the second method for the reconstruction of the participant plane, while the ZDC detectors are used for the estimation of the spectator plane. The signal from both sets of detectors was corrected for detector anisotropies following the procedure described in previous studies~\cite{Arnaldi:1999zz,Acharya:2017fau}. The choice of V0A or V0C used to construct the $q$-vectors and the participant plane follows the analysis configuration and is not expected to affect the extracted CME upper limits.

\subsection{Event shape engineering}

The selection on the shape of the event follows the prescription described in Ref.~\cite{Schukraft:2012ah} 
and is based on the magnitude of the second-order normalized flow vector, $q_2$~\cite{STAR:2002hbo}, given by
\begin{equation}
q_2 = \frac{|\vec{Q}_2|}{\sqrt{M}},
\label{Eq:Q2}
\end{equation}
where $\vec{Q_2}$ is the second-order flow vector whose magnitude is $|\vec{Q}_2| = \sqrt{Q_{2,x}^2 + Q_{2,y}^2}$. The $\vec{Q}_2$ is determined from the azimuthal distribution of the energy deposition measured in the V0C detector with $Q_{2,x}$ and $Q_{2,y}$ being its x- and y-components calculated as
\begin{equation}
Q_{2,x} = \sum_i w_i \cos(2\varphi_i),~~Q_{2,y} = \sum_i w_i \sin(2\varphi_i).
\label{Eq:Q2xy}
\end{equation}
The sum runs over all channels $i$ of the V0C detector ($i = 1-32$), 
$\varphi_i$ is the azimuthal angle of channel $i$, $w_i$ is the amplitude measured in channel 
$i$, and $M = \sum_i w_i$. 
The $q_2$ distribution is divided in percentiles from where ten event-shape classes are defined with the lowest (highest) $q_2$ value corresponding to the 0--10\% (90--100\%) 
range and are studied for each centrality interval. The results for each $q_2$ range correspond to the ``biased'' measurements, while the relevant results without any $q_2$ selection are referred to as ``unbiased''.

The elliptic-flow coefficient is measured using the event-plane method~\cite{Poskanzer:1998yz} 
according to
\begin{equation}
v_2\{\rm EP\} = \langle \langle \cos[2(\varphi_i - \Psi_2)] \rangle \rangle / {\mathcal R}_{2},
\label{Eq:v2EP}
\end{equation}
where the brackets $\langle \langle \cdots \rangle \rangle$ denote 
an average over all events and an average over all particles in all events. In addition, 
$\varphi_i$ is the azimuthal angle of the particle of interest $i$ measured in the TPC, and 
$\psid$, the second order event plane, estimates $\Psi_{\mathrm{PP}}$. It is reconstructed from the azimuthal distribution of the energy deposition measured by the V0A detector as
\begin{equation}
\psid^{\mathrm{V0A}} = \text{atan2}[Q^{\mathrm{V0A}}_{2, y}/ Q^{\mathrm{V0A}}_{2, x}]/2.
\label{Eq:Psi2}
\end{equation}
The event-plane resolution correction factor $\mathcal R_2$ in Eq.~\ref{Eq:v2EP} is calculated from correlations 
between the event planes determined in the TPC, V0C, and V0A detectors~\cite{Poskanzer:1998yz}, evaluated separately for each $q_2$ interval and reaches a maximum value of 0.79 for the 20--30\% centrality class. Non-flow contributions to both $v_2$ and charge-dependent azimuthal correlations are suppressed by the large pseudorapidity separation between the TPC and the V0A ($|\Delta \eta| > 2.0$).

\subsection{Charge-dependent correlations relative to different symmetry planes}
\label{sec:sppp}

This method, proposed in Refs.~\cite{Xu:2017qfs,Voloshin:2018qsm}, relies on the changes in the 
relative strength of the flow induced background and the potential CME signal when charge-dependent 
correlations are measured with respect to $\Psi_{\mathrm{PP}}$ or $\Psi_{\mathrm{SP}}$ planes. In particular, since the background depends on the elliptic-flow 
magnitude which, when calculated from correlations between two particles, is maximum in the participant plane, its contribution to $\Delta \gamma$ will be 
enhanced or reduced when the charge-dependent correlations are measured relative to 
$\Psi_{\mathrm{PP}}$ or $\Psi_{\mathrm{SP}}$, respectively. Inversely, since the CME signal is 
proportional to the value of the magnetic field which is mainly determined by the spectator protons, 
measurements performed relative to $\Psi_{\mathrm{SP}}$ will have larger contribution from it than 
when the same measurements are performed with respect to $\Psi_{\mathrm{PP}}$. One can then form 
the double ratio $(\Delta\gamma/v_2)_{\text{SP}}/(\Delta\gamma/v_2)_{\text{PP}}$ which is expected 
to be larger than unity in the case of non-zero CME. In that case the double ratio can be used to 
estimate the relative contribution of the CME to $\Delta\gamma$, expressed as the fraction \fcme. 
Assuming that both the spectator plane and the plane perpendicular to the magnetic field coincide, the \fcme~can be extracted according to
\begin{equation}
\label{Eq:doubleRatio}
    \frac{(\Delta\gamma/v_2)_{\text{SP}}}{(\Delta\gamma/v_2)_{\text{PP}}}=1+f_{\text{CME}}\left(\frac{\langle v_{2,\text{PP}}^2\rangle}{v^2_{2,\text{SP}}}-1\right).
\end{equation}

In Ref.~\cite{Voloshin:2018qsm}, it was noticed that the ratio $\Delta\gamma/v_2$ can be measured directly as 
\begin{equation}
\label{Eq:SPPPratioEquation}
    (\Delta\gamma/v_2)_{\mathrm{PP,SP}} = \frac{\Delta\langle\cos{(\varphi_\alpha+\varphi_\beta-2\Psi_{\mathrm{PP,SP}})}\rangle}{\langle\cos{(2\varphi_a-2\Psi_{\mathrm{PP,SP}})}\rangle},
\end{equation}
which does not require the knowledge of the reaction-plane resolution, and thus could be measured 
more precisely. In our measurements, the spectator plane is calculated from the combined signal of the 
two ZDC detectors on the A- and C-sides using equation analogous to Eq. \ref{Eq:Q2xy} for a first harmonic~\cite{ALICE:2013xri} to calculate the corresponding Q-vectors $\vec{Q}^{\text{ZDC-A}}$ and $\vec{Q}^{\text{ZDC-C}}$, and $\Psi_{\mathrm{PP}}$ is estimated using the V0C detector. Similarly to the previous method, the $v_2$ is measured using the event-plane method~\cite{Poskanzer:1998yz}.
The large pseudorapidity difference between the TPC, where the particles of interest for both the 
calculation of $\gamma$ and $v_2$ are reconstructed, and the V0C detector in one case or the ZDC 
in the other case, ensures a significant reduction of non-flow contributions to the measurement.

%

\section{Systematic uncertainties}
\label{Sec:Uncertainties}

The systematic uncertainties for both methods are evaluated by commonly varying the selection 
criteria at both the event and track levels for both analyses. All common sources listed below are investigated 
for the biased and unbiased results for same- and opposite-sign charge combinations in $\gamma$, 
as well as for $v_2$ for the ESE method. Similarly, for the double ratio in Eq.~\ref{Eq:SPPPratioEquation}, 
the results of correlations relative to different symmetry planes for each centrality interval are studied separately for the various charge combinations for $\gamma$ 
as well as for $v_2$. This procedure allowed to take into account any correlations between the different components that form the more complex observables (e.g. Eq.~\ref{Eq:doubleRatio}) reported in this article. 
Only statistically significant differences of more than $2\sigma$ between the values of the nominal 
data points and the results from the variations are assigned as systematic uncertainties according 
to Ref.~\cite{Barlow:2002yb}. The contributions from the various independent sources are added in quadrature to obtain the final 
systematic uncertainties on each of the measurements. These are represented with a box around each 
data point, while the statistical uncertainties are denoted by vertical bars in all figures of this 
article.
 
Systematic effects originating from the variation of the acceptance of the detector are investigated by varying 
the range of the position of the reconstructed primary vertex on the z-axis from its default value of 
$|V_z| < 10$~cm down to $|V_z| < 5$~cm. Potential biases coming from the centrality determination are 
studied by replacing the V0 detector amplitude by the total number of hits detected in two SPD layers. Systematic effects originating from the magnetic-field polarity are 
studied by analyzing events recorded under different magnetic field polarities independently. In addition, the 
influence of pileup events to both measurements is estimated by either completely removing the pileup 
requirements or by imposing stricter selections. Effects introduced by the different track selection 
criteria are checked for both analyses by varying the minimum number of TPC space points (i.e., from 70 to 90), the 
minimum number of crossed TPC pad rows and the expected fraction of TPC space points to reconstruct a
track (i.e., from 70 to 120 and from 0.8 to 0.9, respectively), 
and by imposing tighter DCA requirements that reduce the contribution from secondary particles to the 
track sample. Biases from the detector combination used to perform track reconstruction are tested by employing an alternative method, 
referred to as global tracking. This mode combines the TPC and ITS detector signals and incorporates stricter 
selection criteria, including additional hit requirements in the SPD or SDD detectors. In addition, 
for both methods the results for same-sign correlations obtained using the combination of only positive 
or only negative particles are compared and found to be compatible. Finally, 
potential biases introduced by the inaccurate estimation of the reconstruction efficiency are studied 
by either not applying any efficiency corrections or by estimating these corrections after excluding the 
contribution from strange and multi-strange particles. Among these common sources, the dominant contributions for the ESE method are associated with the centrality determination and the alternative track reconstruction configuration (global tracking), whereas for the (PP/SP) method the dominant contribution is discussed below.

In addition to the previous common contributions, the (PP/SP) method includes an extra source related 
to noticeable un-physical off-diagonal correlations $\langle XY \rangle$ ($\langle Q_x^{\text{ZDC-C}}Q_y^{\text{ZDC-A}}\rangle$) and $\langle YX\rangle$ ($\langle Q_y^{\text{ZDC-C}}Q_x^{\text{ZDC-A}}\rangle$) 
components even after the ZDC signal calibration.
These residual non-zero values for $\langle XY\rangle$ and $\langle YX\rangle$ are due to detector and beam-related artefacts. They could arise from a small rotation or shear of the spectator-neutron spot at the ZDC location due to the specific heavy-ion beam optics during the LHC operations (e.g. beam divergence and crossing angle), combined with remaining misalignment of the ZDCs with respect to the interaction region and their response non-uniformities. These effects can lead to an effective mixing of the $X$ and $Y$ components that is not removed by the signal correction procedure~\cite{ALICE:2022xhd}.
This residual correlation affects the values of $\cos{(n\Psi^{\text{ZDC-C,ZDC-A}}_{\text{SP}}})$ and 
$\sin{(n\Psi^{\text{ZDC-C,ZDC-A}}_{\text{SP}}})$. To evaluate the potential impact on $\Delta\gamma_{\text{SP}}$ and 
$v_{2,\text{SP}}$, the approach used in Ref.~\cite{ALICE:2022xhd} is adopted, where the values of these 
observables are scaled by a factor ($F_{\rm ZDC}$) to estimate any residual effects
\begin{equation}
\label{CrossTerm1}
    F_{\rm ZDC}=\sqrt{\frac{\langle Q_x^{\text{ZDC-C}}Q_y^{\text{ZDC-A}}\rangle\langle Q_y^{\text{ZDC-C}}Q_x^{\text{ZDC-A}}\rangle}{\langle Q_x^{\text{ZDC-C}}Q_x^{\text{ZDC-A}}\rangle\langle Q_y^{\text{ZDC-C}}Q_y^{\text{ZDC-A}}\rangle}}.
\end{equation}

For the (PP/SP) method, the event-plane resolution for both $\Psi_{\mathrm{SP}}$ and $\Psi_{\mathrm{PP}}$ cancel out in the 
ratio $(\Delta\gamma/v_2)$ ~\cite{Voloshin:2018qsm}. However, extracting the CME fraction from 
Eq.~\ref{Eq:doubleRatio} requires accounting for the event-plane resolutions of both $\Psi_{\mathrm{PP}}$ 
and $\Psi_{\mathrm{SP}}$ in the ratio $\langle v^2_{2,\text{PP}}\rangle/v^2_{2,\text{SP}}$. Since 
the resolution of the spectator plane could not be directly determined using the ALICE ZDC detector 
for the Pb--Pb events at $\snn=5.02$~TeV~recorded in 2018, we assume that the ratio 
$v_{2,\text{SP}}/v_{2,\text{PP}}$ is independent of the collision energy
\begin{equation}
    \frac{v_{2,\text{SP}}(\text{5.02\,TeV})}{v_{2,\text{PP}}(\text{5.02\,TeV})}=\frac{v_{2,\text{SP}}(\text{2.76\,TeV})}{v_{2,\text{PP}}(\text{2.76\,TeV})},
\end{equation}
where the spectator-plane resolution for the Pb--Pb events at $\snn=2.76$~TeV recorded in 2010 was 
successfully determined~\cite{ALICE:2022xhd}. The ratio primarily reflects the strength of flow fluctuations, i.e., the decorrelation between the reaction plane and the participant plane due to initial‐geometry fluctuations. These fluctuations are determined by the transverse nucleon distribution and are not expected to vary significantly with LHC energies~\cite{Broniowski:2007ft, Drescher:2007ax}. In addition, none of the common sources of systematic uncertainty results in a statistically 
significant deviation greater than 2$\sigma$ in the double ratio. This shows that the double ratio is 
particularly effective in canceling out systematic uncertainties that are unrelated to the spectator-plane 
reconstruction. The only contribution to the systematic uncertainty in the double ratio originates from 
the biases in the estimation of the spectator plane, whose magnitude is comparable to the statistical uncertainty.

\section{Results}
\label{Sec:Results}

\subsection{Event shape engineering}

\begin{figure}[tp]
 \centering
 \includegraphics[keepaspectratio, width=0.65\columnwidth]{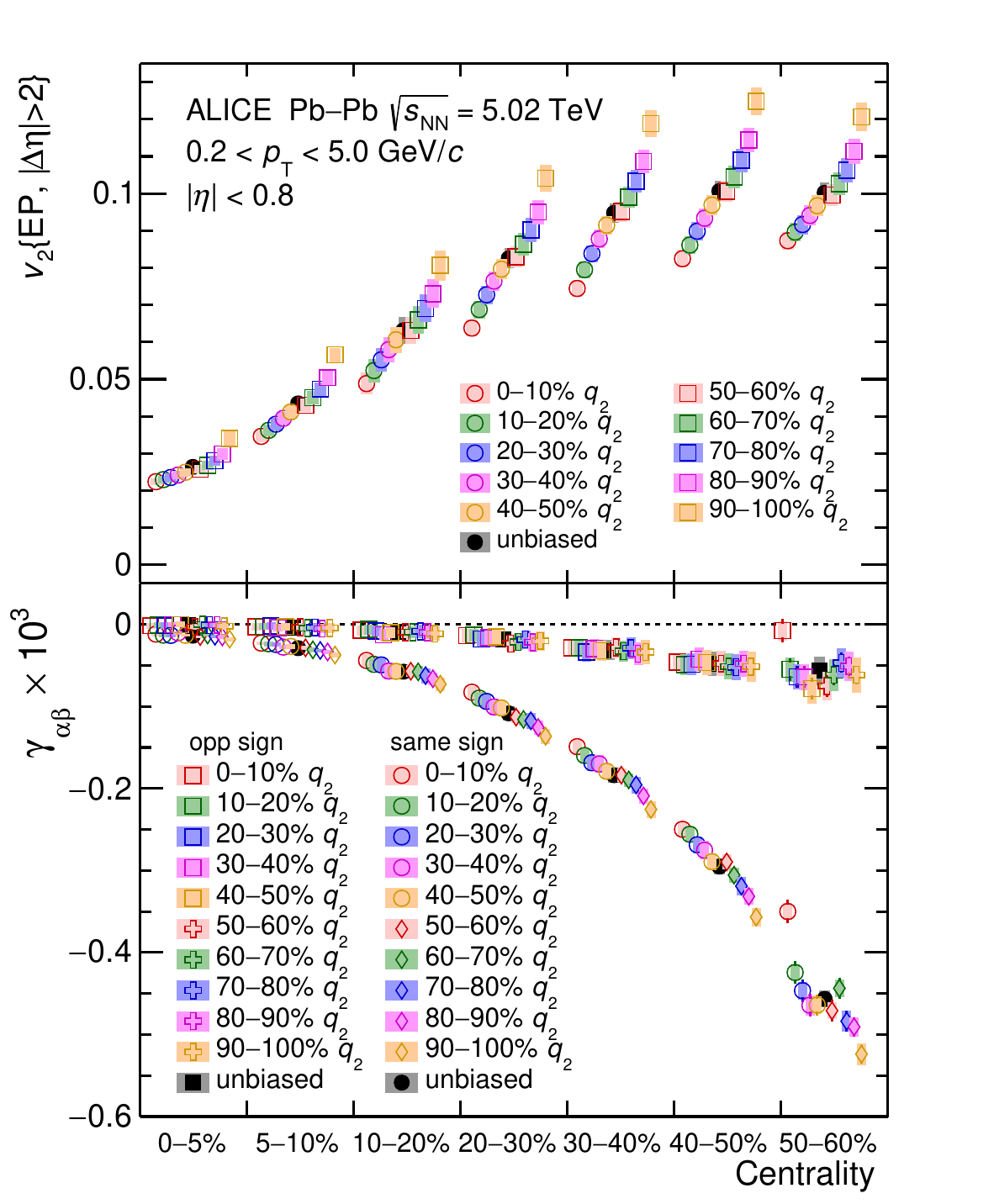}
 \caption{Centrality dependence of inclusive charged particle $v_2$ (top) and of 
 $\gc$ for pairs of particles with same and opposite sign (bottom) for shape-selected and unbiased 
 events. The $q_2$ determined in the V0C is used for event selection. Statistical and systematic 
 uncertainties are represented by vertical lines and shaded boxes, respectively.}
 \label{fig:v2_gamma_ESE}
\end{figure}

The top panel of Fig.~\ref{fig:v2_gamma_ESE} presents the centrality dependence of $v_2$ for inclusive 
charged particles measured in $0.2<\pt<5.0$ GeV/$c$ and $|\eta| < 0.8$ for shape-selected and unbiased 
samples. The event shape-selected results differ from the average by up to 30\%. These differences 
vary strongly with centrality as the sensitivity of the event selection depends on multiplicity and the 
magnitude of $v_2$~\cite{ALICE:2015lib}. Furthermore, $v_2$ scales approximately linearly with 
eccentricity~\cite{Gardim:2011xv} verifying that with the ESE method events with different eccentricity 
(i.e., initial spatial anisotropy) and, therefore, with different background contribution are 
experimentally selected in a given centrality interval.

The bottom panel of Fig.~\ref{fig:v2_gamma_ESE} shows the centrality dependence of $\gc$ for same- and 
opposite-sign pairs. Since the results for pairs of particles with only positive charges are consistent 
with those for negative charges, they are averaged and denoted as same-sign pairs in all figures and 
for both methods. Similarly to what is reported in previous measurements~\cite{Acharya:2017fau}, a clear dependence 
on the charge-sign combination is exhibited by $\gc$ with the opposite-sign pair correlations being 
weaker than the same-sign pair ones for both unbiased and shape-selected events. The magnitude of the 
same-sign pair correlations decreases from central to peripheral collisions and depends on the 
event-shape selection (i.e., $v_2$) in a given centrality interval. The magnitude of opposite-sign 
pair correlations is close to zero with a weak (if any) dependence on $v_2$ within a given centrality 
interval.

\begin{figure}[htbp]
 \centering
 \includegraphics[keepaspectratio, width=0.65\columnwidth]{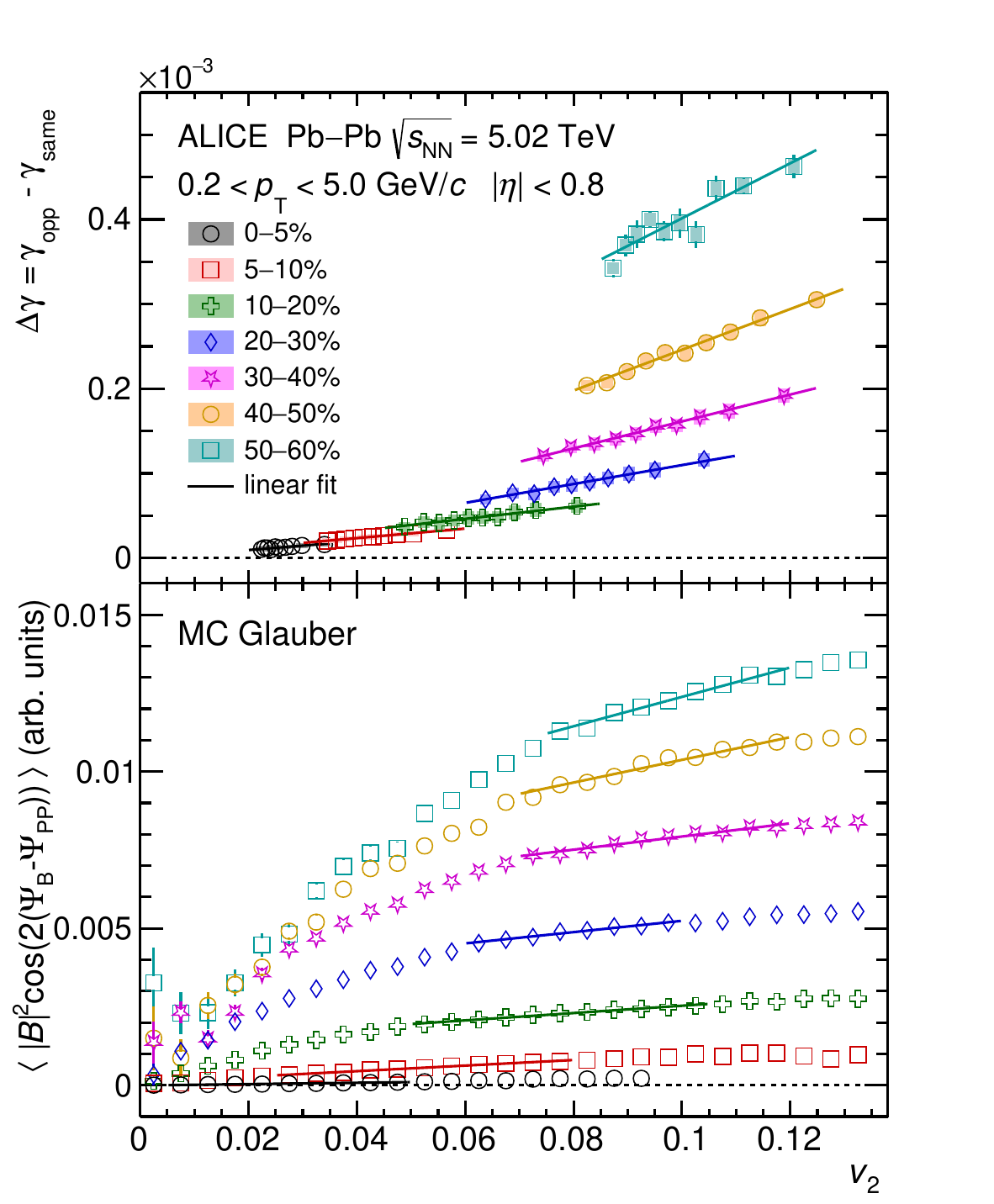}
 \caption{The dependence of $\Delta\gc$ (top) and of the expected CME signal 
 from a MC Glauber simulation~\cite{Miller:2007ri} (bottom) on $v_2$ together with linear fits (solid lines) for 
 various centrality intervals. For the model, no event shape selection is performed and the range of the 
 fit is based on the $v_2$ variation observed in data within each centrality class. Statistical and systematic 
 uncertainties are represented by vertical lines and shaded boxes, respectively.}
 \label{fig:v2_gammaDif_CMEGlb_ESE}
\end{figure}

In a background-dominated scenario, the charge-dependent effects, reflected in the charge-dependent 
differences $\Delta \gamma$, would be proportional 
to $v_2$ for all centrality intervals. The upper panel of Fig.~\ref{fig:v2_gammaDif_CMEGlb_ESE} presents 
the dependence of $\Delta \gamma$ on the main background gauge, i.e., $v_2$, for various  
ESE selections within each centrality interval. It is seen that $\Delta \gamma$ is positive and exhibits a 
clear linear dependence on $v_2$, also illustrated by the linear functions fitted on each set of data 
points. In addition, $\Delta \gamma$ has a clear centrality dependence, with the values progressively 
increasing when moving from central to peripheral events. This is expected considering dilution effects~\cite{ALICE:2022wpn} 
introduced by the higher multiplicity in central compared with peripheral events.   

The bottom panel of Fig.~\ref{fig:v2_gammaDif_CMEGlb_ESE} shows the $v_2$ dependence of the expected 
CME signal from a MC Glauber simulation~\cite{Miller:2007ri} for various centrality intervals. The CME 
signal contribution to $\Delta \gamma$ is assumed to be proportional to $\langle |B|^2 \cos[2(\Psi_{\rm B} - \Psi_{\rm PP})] \rangle$, where 
$\Psi_{\rm B}$ and $|B|$ are the direction and magnitude of the magnetic field, respectively. The 
magnetic field is evaluated at the geometrical center of the overlap region from the number of spectator protons with the proper time 
$\tau=0.1$~fm/$c$ according to the recipe discussed in Ref.~\cite{Kharzeev:2007jp}. The elliptic flow calculated from the model reproduces fairly well the measured integrated $v_2$~\cite{Acharya:2018ihu} under the assumption that 
it scales linearly with eccentricity calculated from the distribution of participant nucleons. Centrality 
is determined as in Ref.~\cite{Acharya:2017fau} from the multiplicity of charged particles generated 
in the V0 acceptance using the parameters reported in Ref.~\cite{ALICE:2013hur}. A strong dependence 
on $v_2$ is found for all centrality intervals which is also reproduced by a MC simulation using 
T$_{\rm R}$ENTo initial conditions~\cite{Moreland:2014oya}. 

\begin{figure}[tp]
 \centering
 \includegraphics[keepaspectratio, width=0.65\columnwidth]{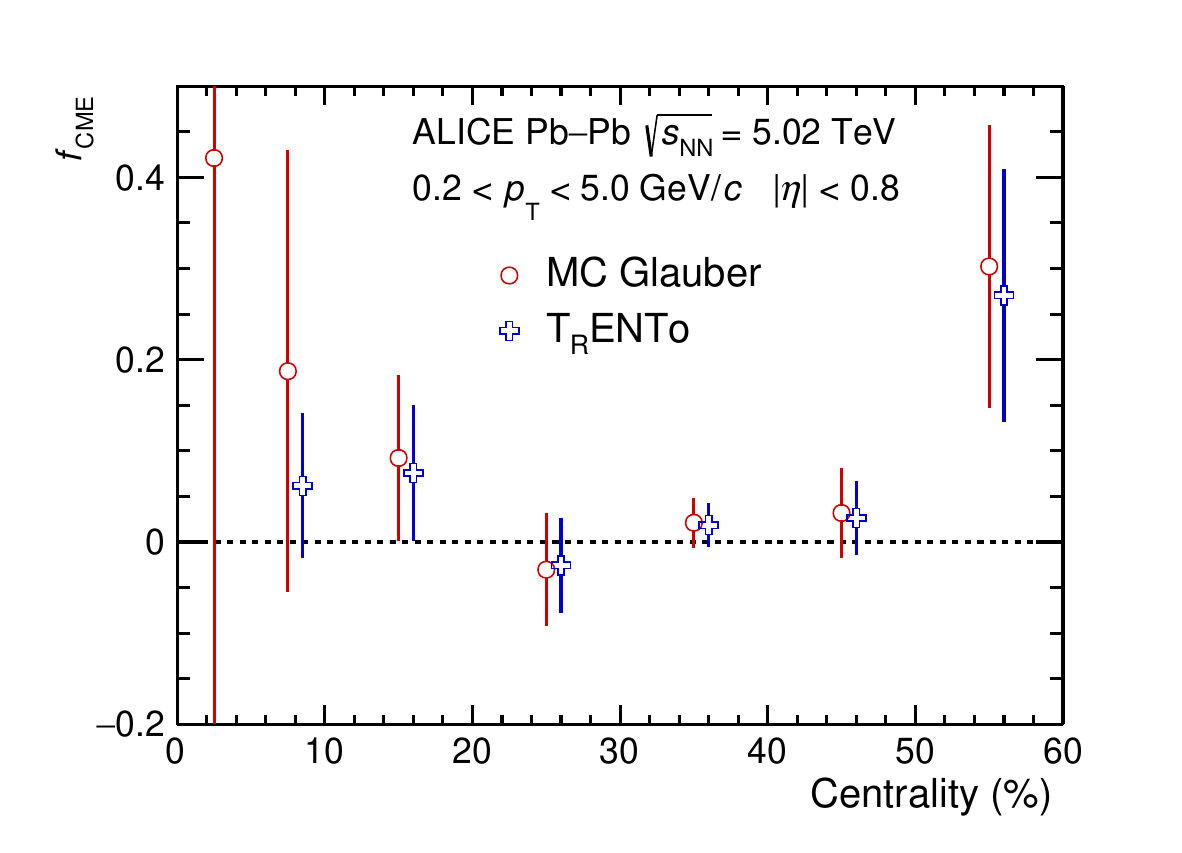}
 \caption{Centrality dependence of the CME fraction
   extracted from the slope parameter of fits to data and
   MC Glauber~\cite{Miller:2007ri} and T$_{\rm R}$ENTo~\cite{Moreland:2014oya} models (see text for details). 
   The T$_{\rm R}$ENTo points are slightly shifted along the horizontal axis for better visibility. 
   Only statistical uncertainties are shown.}
\label{fig:fCME_ESE}
\end{figure}

Following the approach described in Ref.~\cite{Acharya:2017fau}, the $v_2$ dependence of $\Delta\gamma$ 
and that of the expected CME signal are fitted with linear functions in the range that matches the one 
measured in collision data for each centrality interval presented in the upper panel of 
Fig.~\ref{fig:v2_gamma_ESE}. This range was varied, and no significant difference was found in the results described below. The slopes from data and MC models 
are used to calculate the CME fraction to the charge dependence of \gc{}, denoted as $f_{\rm CME}$. This relies on expressing the extracted slope parameter as a linear superposition of a background contribution, which scales proportionally with $v_{2}$ and therefore has $p_{1}=1$, and a CME contribution, whose slope is given by the MC expectation $p_{1,\mathrm{MC}}$, according to

\begin{equation}
f_{\mathrm{CME}} \times p_{1,\mathrm{MC}} + (1 - f_{\mathrm{CME}}) \times 1 = p_{1,\mathrm{data}},
\end{equation}

\noindent from where one gets

\begin{equation}
f_{\mathrm{CME}} = \frac{p_{1,\mathrm{data}} - 1}{p_{1,\mathrm{MC}} - 1}.
\end{equation}

The extracted $f_{\mathrm{CME}}$ therefore quantifies the fraction of the charge-dependent correlator $\Delta \gamma$ that is attributed to the CME contribution under the assumption that the remaining part originates from background correlations proportional to $v_{2}$.

Figure~\ref{fig:fCME_ESE} shows the centrality dependence of $f_{\rm CME}$ for the MC Glauber and 
T$_{\rm R}$ENTo models. The CME fraction cannot be precisely determined for the 0--5\% centrality 
interval where models have difficulties in reproducing the experimental data~\cite{Acharya:2018ihu}. 
The values of $f_{\rm CME}$ are compatible with zero for most of the investigated centrality classes. 
The reduced sensitivity of the event selection (e.g. with the ESE) contributes to the large values obtained in 
the 50--60\% centrality interval. A centrality-dependent extraction of the $f_{\rm CME}$ requires a knowledge
about the magnetic-field evolution and of the chiral imbalance as a
function of centrality, which are currently not constrained neither
experimentally nor theoretically.
For these reasons, an averaged limit for the 5–60\% centrality interval
is obtained by fitting the data points with a constant function is 
$f_{\rm CME} = 0.028 \pm 0.021$ and $f_{\rm CME} = 0.025 \pm 0.018$ for MC Glauber and T$_{\rm R}$ENTo 
initial conditions, respectively. These values correspond to upper limits for the CME fraction in $\Delta \gamma$ of 7\% and 
6\% at 95\% confidence level for the 5--60\% centrality interval.

\subsection{Charge-dependent correlations relative to different symmetry planes (PP/SP)}

\begin{figure}[tp]
\centering
     \includegraphics[width=0.65\linewidth]{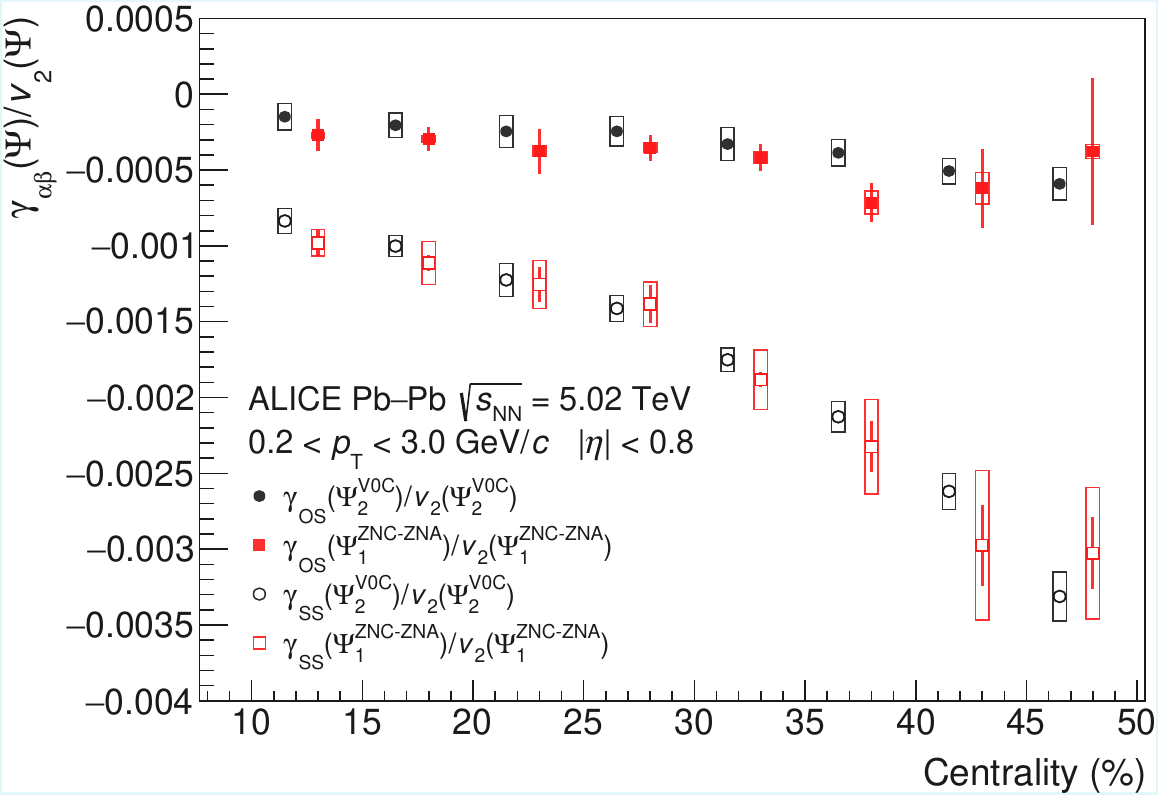}
\caption{Centrality dependence of $\gamma_{\alpha\beta}(\Psi_{\text{PP}})/ v_2(\Psi_{\text{PP}})$ 
(black) and $\gamma_{\alpha\beta}(\Psi_{\text{SP}})/v_2(\Psi_{\text{SP}})$ (red), where pairs of 
particles in $\gamma_{\alpha\beta}$ are either of the same charge or the opposite charge in Pb--Pb collisions 
at $\snn = 5.02$~TeV. Filled markers correspond to OS correlations, open markers to SS correlations, with vertical bars representing statistical uncertainties and boxes representing systematic uncertainties.}
\label{gammaDivV2}
\end{figure}

Figure~\ref{gammaDivV2} presents the centrality dependence of $\gamma_{\alpha\beta}(\Psi_{\text{PP}})/v_2(\Psi_{\text{PP}})$ and $\gamma_{\alpha\beta}(\Psi_{\text{SP}})/v_2(\Psi_{\text{SP}})$, where pairs 
of particles in $\gamma_{\alpha\beta}$ are either of the same or opposite charge.  
A significant charge dependence that develops as a function of centrality is observed in both categories 
of results. Moreover, it is seen that the two results do not exhibit any significant differences for 
a given charge combination.

\begin{figure}[H]
\centering
    \includegraphics[width=0.65\linewidth]{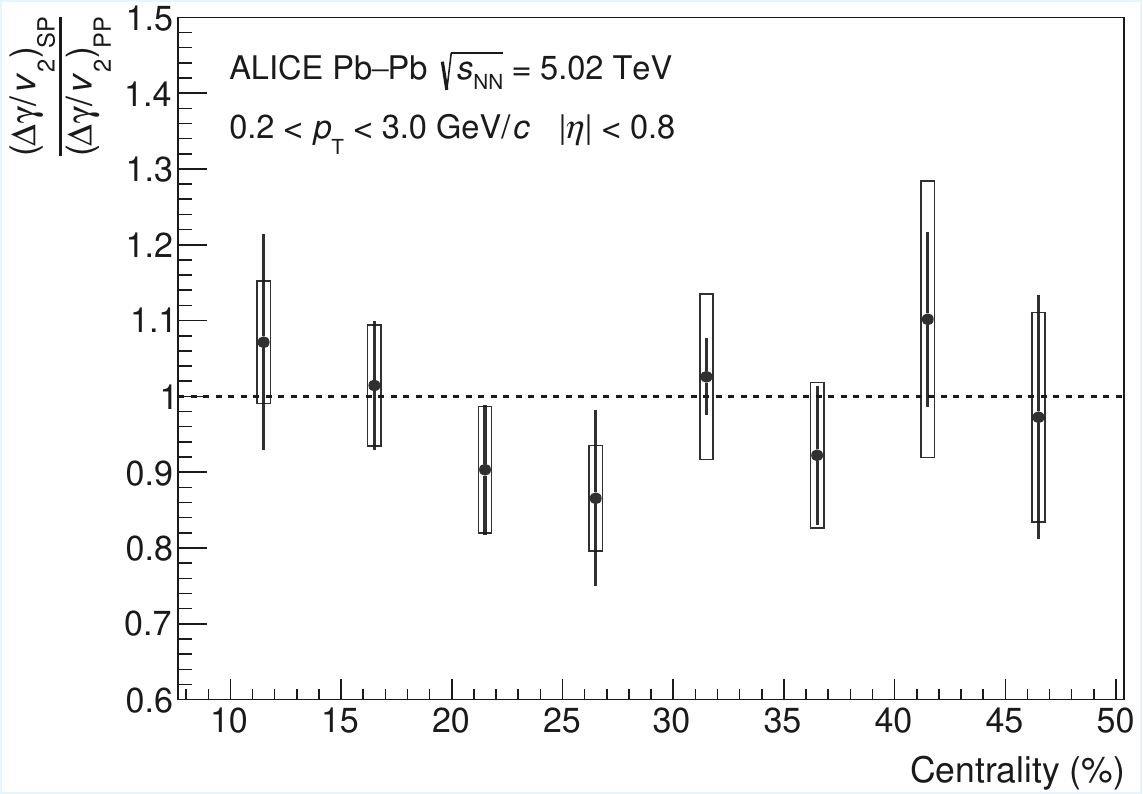}
    \caption{The double ratio of $\Delta \gamma/v_2$ in Eq.~\ref{Eq:doubleRatio} calculated with the 
    spectator and participant planes as a function of centrality in Pb--Pb collisions at $\snn = 5.02$~TeV. The vertical bars and boxes represent statistical and systematic uncertainties, respectively.}
    \label{DoubleRatioZDCvsV0C}
\end{figure}

\begin{figure}[H]
\centering
	\centering
     \includegraphics[width=0.7\linewidth]{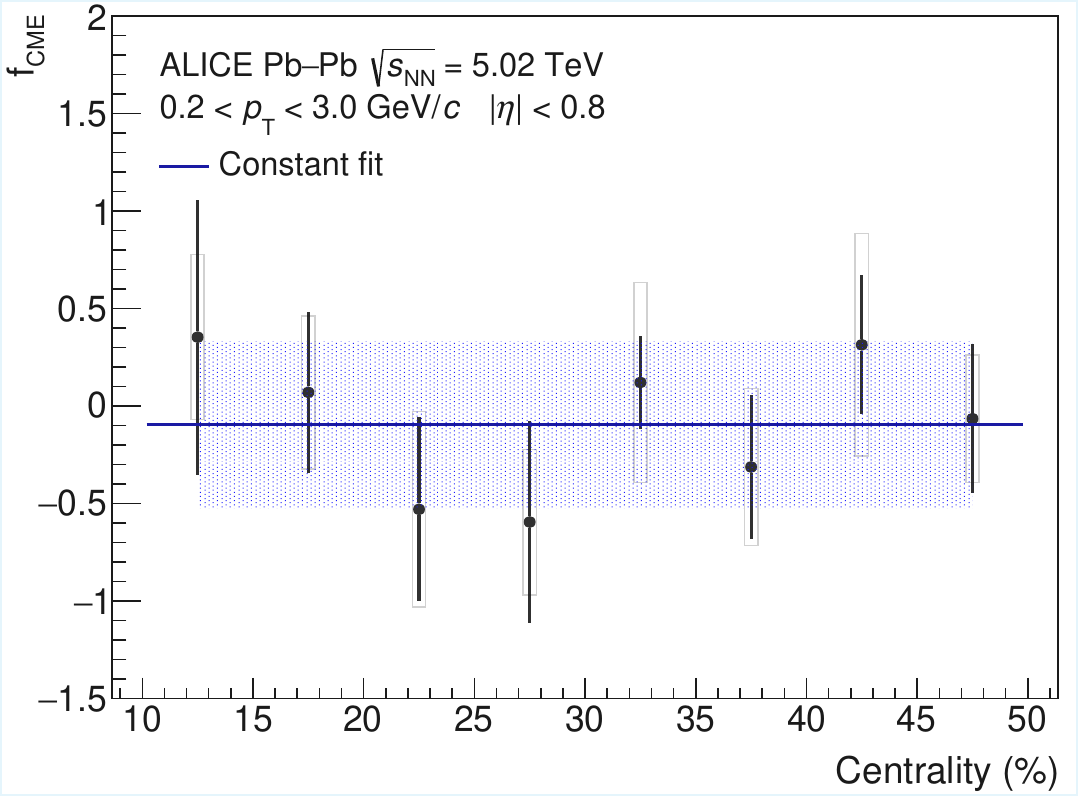}
\caption{The centrality dependence of $f_{\text{CME}}$ extracted from correlations measured relative 
to the spectator and participants planes according to Eq.~\ref{Eq:doubleRatio} in Pb--Pb collisions 
at $\snn = 5.02$~TeV. Statistical uncertainties are shown as vertical bars, and systematic uncertainties are shown as boxes. A constant fit is shown along with a 95\% confidence level band.}
\label{fCMEPlot}
\end{figure}

To quantify the difference, the result of the double ratios in Eq.~\ref{Eq:SPPPratioEquation} 
is shown in Fig.~\ref{DoubleRatioZDCvsV0C}. The double ratios in the centrality 
range of 10-30\% are measured with the minimum-bias trigger, while those in the 30-50\% centrality 
range are measured with the dedicated trigger that enhances events in this centrality interval. The double ratio is compatible with unity within the statistical and systematic 
uncertainties throughout the centrality range 10--50\%.

Figure~\ref{fCMEPlot} shows the extracted $f_{\text{CME}}$ as a function of centrality using the procedure described in Sec. \ref{sec:sppp} and
Eq.~\ref{Eq:doubleRatio}. A constant line fit to $f_{\text{CME}}$ considering the total uncertainty 
that includes both statistical and systematic uncertainties in the centrality range of 10--50\% yields a value of $-0.096 \pm 0.214$. This translates to an upper limit for the CME fraction in $\Delta \gamma$
of 33\% at a 95\% confidence level.

\subsection{ALICE summary of upper limits for the CME fraction in \texorpdfstring{$\Delta \gamma$}{Delta gamma}}

\begin{figure}[H]
 \centering
 \includegraphics[keepaspectratio, width=0.75\columnwidth]{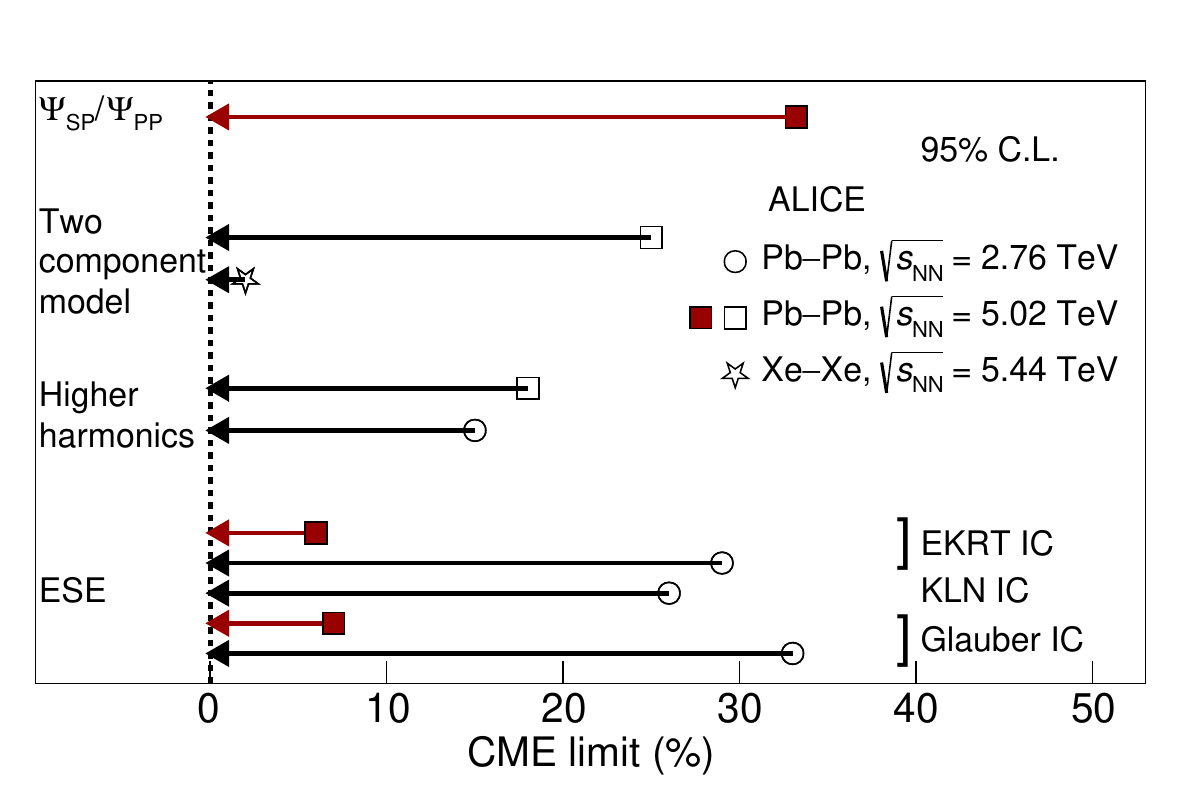}
 \caption{Summary of the results for the CME limit at 95$\%$ confidence 
 level (C.L.) obtained from different analyses for different colliding systems at various energies 
 at the LHC integrated over centralities from ALICE. The previous measurements are taken from Refs.~\cite{ALICE:2020siw, Acharya:2017fau, ALICE:2022ljz} and are represented with the open markers. }
\label{fig:fCME_Limit}
\end{figure}

Figure~\ref{fig:fCME_Limit} presents a summary of the upper limits reported by ALICE for the CME signal 
contribution to $\Delta \gamma$. These values are obtained from different analyses performed at various LHC energies 
and colliding systems integrated over centralities in Run 1 and Run 2. These include the data points from previous ESE 
studies~\cite{Acharya:2017fau} incorporating results from three initial-state models, higher harmonics 
at two different LHC energies~\cite{ALICE:2020siw}, as well as from the comparison of correlations 
in Xe--Xe and Pb--Pb collisions using a two-component model~\cite{ALICE:2022ljz}. The initial ESE 
results provided the first upper limits at the LHC of around 26--33\% at 95\% confidence 
level~\cite{Acharya:2017fau}, while the ones from the higher harmonic study that rely on specific 
assumptions discussed in detail in Ref.~\cite{ALICE:2020siw} reported similar limits of around 
15--18$\%$ at 95$\%$ confidence level. Finally, in Ref.~\cite{ALICE:2022ljz} ALICE reported an upper 
limit of around 2\% and 25$\%$ at 95$\%$ confidence level for the 0--70$\%$ centrality interval in 
Xe--Xe and Pb--Pb collisions, respectively. The main reason for the low upper limit reported in the Xe--Xe system is related to the significantly lower magnetic field value this system creates relative to the Pb--Pb one. The measurements reported in this article add significant 
information to the existing set of results.

\section{Summary}
\label{Sec:Summary}

In this article, the results from the analysis of Pb--Pb collisions at $\snn = 5.02$~TeV collected 
with the ALICE detector using two independent methods are reported. The charge-dependent correlations 
are studied using the correlator $\gc$ with the ESE technique and by measuring 
correlations relative to the participant and spectator planes. These two methods have distinctively 
different sensitivities to the background originating from local charge conservation effects embedded 
in an anisotropically expanding system and the potential CME signal. Both methods reported significant 
charge dependent effects which, however, are compatible with a dominant background contribution. In 
both cases upper limits for the CME fraction in $\Delta \gamma$ of 7\% and 6\% at 95\% confidence level for the 5--60\% centrality interval 
and 33\% at a 95\% confidence level for the 10--50\% centrality interval, respectively, are reported. 
These results, while having different uncertainties due to experimental techniques, are equally important for the CME search and complement the limits obtained from previous relevant measurements at LHC energies. With the significantly larger data sample from Run 3 and Run 4, analyses based on ESE and correlations relative to different symmetry planes will achieve higher precision. In addition, new measurements involving identified particle species~\cite{Citron:2018lsq} and the use of the ZDC as a sensitive probe of the magnetic field~\cite{Christakoglou:2023hgk} will provide complementary avenues to further advance in the CME search.


\newenvironment{acknowledgement}{\relax}{\relax}
\begin{acknowledgement}
\section*{Acknowledgements}

The ALICE Collaboration would like to thank all its engineers and technicians for their invaluable contributions to the construction of the experiment and the CERN accelerator teams for the outstanding performance of the LHC complex.
The ALICE Collaboration gratefully acknowledges the resources and support provided by all Grid centres and the Worldwide LHC Computing Grid (WLCG) collaboration.
The ALICE Collaboration acknowledges the following funding agencies for their support in building and running the ALICE detector:
A. I. Alikhanyan National Science Laboratory (Yerevan Physics Institute) Foundation (ANSL), State Committee of Science and World Federation of Scientists (WFS), Armenia;
Austrian Academy of Sciences, Austrian Science Fund (FWF): [M 2467-N36] and Nationalstiftung f\"{u}r Forschung, Technologie und Entwicklung, Austria;
Ministry of Communications and High Technologies, National Nuclear Research Center, Azerbaijan;
Rede Nacional de Física de Altas Energias (Renafae), Financiadora de Estudos e Projetos (Finep), Funda\c{c}\~{a}o de Amparo \`{a} Pesquisa do Estado de S\~{a}o Paulo (FAPESP) and The Sao Paulo Research Foundation  (FAPESP), Brazil;
Bulgarian Ministry of Education and Science, within the National Roadmap for Research Infrastructures 2020-2027 (object CERN), Bulgaria;
Ministry of Education of China (MOEC) , Ministry of Science \& Technology of China (MSTC) and National Natural Science Foundation of China (NSFC), China;
Ministry of Science and Education and Croatian Science Foundation, Croatia;
Centro de Aplicaciones Tecnol\'{o}gicas y Desarrollo Nuclear (CEADEN), Cubaenerg\'{\i}a, Cuba;
Ministry of Education, Youth and Sports of the Czech Republic, Czech Republic;
The Danish Council for Independent Research | Natural Sciences, the VILLUM FONDEN and Danish National Research Foundation (DNRF), Denmark;
Helsinki Institute of Physics (HIP), Finland;
Commissariat \`{a} l'Energie Atomique (CEA) and Institut National de Physique Nucl\'{e}aire et de Physique des Particules (IN2P3) and Centre National de la Recherche Scientifique (CNRS), France;
Bundesministerium f\"{u}r Forschung, Technologie und Raumfahrt (BMFTR) and GSI Helmholtzzentrum f\"{u}r Schwerionenforschung GmbH, Germany;
National Research, Development and Innovation Office, Hungary;
Department of Atomic Energy Government of India (DAE), Department of Science and Technology, Government of India (DST), University Grants Commission, Government of India (UGC) and Council of Scientific and Industrial Research (CSIR), India;
National Research and Innovation Agency - BRIN, Indonesia;
Istituto Nazionale di Fisica Nucleare (INFN), Italy;
Japanese Ministry of Education, Culture, Sports, Science and Technology (MEXT) and Japan Society for the Promotion of Science (JSPS) KAKENHI, Japan;
Consejo Nacional de Ciencia (CONACYT) y Tecnolog\'{i}a, through Fondo de Cooperaci\'{o}n Internacional en Ciencia y Tecnolog\'{i}a (FONCICYT) and Direcci\'{o}n General de Asuntos del Personal Academico (DGAPA), Mexico;
Nederlandse Organisatie voor Wetenschappelijk Onderzoek (NWO), Netherlands;
The Research Council of Norway, Norway;
Pontificia Universidad Cat\'{o}lica del Per\'{u}, Peru;
Ministry of Science and Higher Education, National Science Centre and WUT ID-UB, Poland;
Korea Institute of Science and Technology Information and National Research Foundation of Korea (NRF), Republic of Korea;
Ministry of Education and Scientific Research, Institute of Atomic Physics, Ministry of Research and Innovation and Institute of Atomic Physics and Universitatea Nationala de Stiinta si Tehnologie Politehnica Bucuresti, Romania;
Ministerstvo skolstva, vyskumu, vyvoja a mladeze SR, Slovakia;
National Research Foundation of South Africa, South Africa;
Swedish Research Council (VR) and Knut \& Alice Wallenberg Foundation (KAW), Sweden;
European Organization for Nuclear Research, Switzerland;
Suranaree University of Technology (SUT), National Science and Technology Development Agency (NSTDA) and National Science, Research and Innovation Fund (NSRF via PMU-B B05F650021), Thailand;
Turkish Energy, Nuclear and Mineral Research Agency (TENMAK), Turkey;
National Academy of  Sciences of Ukraine, Ukraine;
Science and Technology Facilities Council (STFC), United Kingdom;
National Science Foundation of the United States of America (NSF) and United States Department of Energy, Office of Nuclear Physics (DOE NP), United States of America.
In addition, individual groups or members have received support from:
Czech Science Foundation (grant no. 23-07499S), Czech Republic;
FORTE project, reg.\ no.\ CZ.02.01.01/00/22\_008/0004632, Czech Republic, co-funded by the European Union, Czech Republic;
European Research Council (grant nos. 101220549, 950692), European Union;
Deutsche Forschungs Gemeinschaft (DFG, German Research Foundation) ``Neutrinos and Dark Matter in Astro- and Particle Physics'' (grant no. SFB 1258), Germany;
CONVECS project, CUP C97H23001700002 FESR 2021-2027 program, Italy;
FAIR - Future Artificial Intelligence Research, funded by the NextGenerationEU program (Italy).

\end{acknowledgement}

\bibliographystyle{utphys}   
\bibliography{bibliography}

\newpage
\appendix

%
%

\section{The ALICE Collaboration}
\label{app:collab}
\begin{flushleft} 
\small

D.A.H.~Abdallah\,\orcidlink{0000-0003-4768-2718}\,$^{\rm 134}$, 
I.J.~Abualrob\,\orcidlink{0009-0005-3519-5631}\,$^{\rm 112}$, 
S.~Acharya\,\orcidlink{0000-0002-9213-5329}\,$^{\rm 49}$, 
K.~Agarwal\,\orcidlink{0000-0001-5781-3393}\,$^{\rm II,}$$^{\rm 23}$, 
G.~Aglieri Rinella\,\orcidlink{0000-0002-9611-3696}\,$^{\rm 32}$, 
L.~Aglietta\,\orcidlink{0009-0003-0763-6802}\,$^{\rm 24}$, 
N.~Agrawal\,\orcidlink{0000-0003-0348-9836}\,$^{\rm 25}$, 
Z.~Ahammed\,\orcidlink{0000-0001-5241-7412}\,$^{\rm 132}$, 
S.~Ahmad\,\orcidlink{0000-0003-0497-5705}\,$^{\rm 15}$, 
I.~Ahuja\,\orcidlink{0000-0002-4417-1392}\,$^{\rm 36}$
% ...

\section*{Affiliation Notes}

$^{\rm I}$ Deceased\\
$^{\rm II}$ Also at: INFN Trieste\\
$^{\rm III}$ Also at: Fondazione Bruno Kessler (FBK), Trento, Italy\\
$^{\rm IV}$ Also at: Max-Planck-Institut fur Physik, Munich, Germany\\
$^{\rm V}$ Also at: Czech Technical University in Prague (CZ)\\
$^{\rm VI}$ Also at: Instituto de Fisica da Universidade de Sao Paulo\\
$^{\rm VII}$ Also at: Dipartimento DET del Politecnico di Torino, Turin, Italy\\
$^{\rm VIII}$ Also at: Department of Applied Physics, Aligarh Muslim University, Aligarh, India\\
$^{\rm IX}$ Also at: Institute of Theoretical Physics, University of Wroclaw, Poland\\
$^{\rm X}$ Also at: Facultad de Ciencias, Universidad Nacional Aut\'{o}noma de M\'{e}xico, Mexico City, Mexico\\
$^{\rm XI}$ Also at: University College of Dublin, Dublin, Ireland\\

\section*{Collaboration Institutes}

$^{1}$ A.I. Alikhanyan National Science Laboratory (Yerevan Physics Institute) Foundation, Yerevan, Armenia\\
$^{2}$ AGH University of Krakow, Cracow, Poland\\
$^{3}$ Bogolyubov Institute for Theoretical Physics, National Academy of Sciences of Ukraine, Kyiv, Ukraine\\
$^{4}$ Bose Institute, Department of Physics  and Centre for Astroparticle Physics and Space Science (CAPSS), Kolkata, India\\
$^{5}$ California Polytechnic State University, San Luis Obispo, California, United States\\
$^{6}$ Central China Normal University, Wuhan, China\\
$^{7}$ Centro de Aplicaciones Tecnol\'{o}gicas y Desarrollo Nuclear (CEADEN), Havana, Cuba\\
$^{8}$ Centro de Investigaci\'{o}n y de Estudios Avanzados (CINVESTAV), Mexico City and M\'{e}rida, Mexico\\
$^{9}$ Chicago State University, Chicago, Illinois, United States\\
$^{10}$ China Nuclear Data Center, China Institute of Atomic Energy, Beijing, China\\
$^{11}$ China University of Geosciences, Wuhan, China\\
$^{12}$ Chungbuk National University, Cheongju, Republic of Korea\\
$^{13}$ Comenius University Bratislava, Faculty of Mathematics, Physics and Informatics, Bratislava, Slovak Republic\\
$^{14}$ Creighton University, Omaha, Nebraska, United States\\
$^{15}$ Department of Physics, Aligarh Muslim University, Aligarh, India\\
$^{16}$ Department of Physics, Pusan National University, Pusan, Republic of Korea\\
$^{17}$ Department of Physics, Sejong University, Seoul, Republic of Korea\\
$^{18}$ Department of Physics, University of California, Berkeley, California, United States\\
$^{19}$ Department of Physics, University of Oslo, Oslo, Norway\\
$^{20}$ Department of Physics and Technology, University of Bergen, Bergen, Norway\\
$^{21}$ Dipartimento di Fisica, Universit\`{a} di Pavia, Pavia, Italy\\
$^{22}$ Dipartimento di Fisica dell'Universit\`{a} and Sezione INFN, Cagliari, Italy\\
$^{23}$ Dipartimento di Fisica dell'Universit\`{a} and Sezione INFN, Trieste, Italy\\
$^{24}$ Dipartimento di Fisica dell'Universit\`{a} and Sezione INFN, Turin, Italy\\
$^{25}$ Dipartimento di Fisica e Astronomia dell'Universit\`{a} and Sezione INFN, Bologna, Italy\\
$^{26}$ Dipartimento di Fisica e Astronomia dell'Universit\`{a} and Sezione INFN, Catania, Italy\\
$^{27}$ Dipartimento di Fisica e Astronomia dell'Universit\`{a} and Sezione INFN, Padova, Italy\\
$^{28}$ Dipartimento di Fisica `E.R.~Caianiello' dell'Universit\`{a} and Gruppo Collegato INFN, Salerno, Italy\\
$^{29}$ Dipartimento DISAT del Politecnico and Sezione INFN, Turin, Italy\\
$^{30}$ Dipartimento di Scienze MIFT, Universit\`{a} di Messina, Messina, Italy\\
$^{31}$ Dipartimento Interateneo di Fisica `M.~Merlin' and Sezione INFN, Bari, Italy\\
$^{32}$ European Organization for Nuclear Research (CERN), Geneva, Switzerland\\
$^{33}$ Faculty of Electrical Engineering, Mechanical Engineering and Naval Architecture, University of Split, Split, Croatia\\
$^{34}$ Faculty of Nuclear Sciences and Physical Engineering, Czech Technical University in Prague, Prague, Czech Republic\\
$^{35}$ Faculty of Physics, Sofia University, Sofia, Bulgaria\\
$^{36}$ Faculty of Science, P.J.~\v{S}af\'{a}rik University, Ko\v{s}ice, Slovak Republic\\
$^{37}$ Faculty of Technology, Environmental and Social Sciences, Bergen, Norway\\
$^{38}$ Frankfurt Institute for Advanced Studies, Johann Wolfgang Goethe-Universit\"{a}t Frankfurt, Frankfurt, Germany\\
$^{39}$ Fudan University, Shanghai, China\\
$^{40}$ Gauhati University, Department of Physics, Guwahati, India\\
$^{41}$ Helmholtz-Institut f\"{u}r Strahlen- und Kernphysik, Rheinische Friedrich-Wilhelms-Universit\"{a}t Bonn, Bonn, Germany\\
$^{42}$ Helsinki Institute of Physics (HIP), Helsinki, Finland\\
$^{43}$ High Energy Physics Group,  Universidad Aut\'{o}noma de Puebla, Puebla, Mexico\\
$^{44}$ Horia Hulubei National Institute of Physics and Nuclear Engineering, Bucharest, Romania\\
$^{45}$ HUN-REN Wigner Research Centre for Physics, Budapest, Hungary\\
$^{46}$ Indian Institute of Technology Bombay (IIT), Mumbai, India\\
$^{47}$ Indian Institute of Technology Indore, Indore, India\\
$^{48}$ INFN, Laboratori Nazionali di Frascati, Frascati, Italy\\
$^{49}$ INFN, Sezione di Bari, Bari, Italy\\
$^{50}$ INFN, Sezione di Bologna, Bologna, Italy\\
$^{51}$ INFN, Sezione di Cagliari, Cagliari, Italy\\
$^{52}$ INFN, Sezione di Catania, Catania, Italy\\
$^{53}$ INFN, Sezione di Padova, Padova, Italy\\
$^{54}$ INFN, Sezione di Pavia, Pavia, Italy\\
$^{55}$ INFN, Sezione di Torino, Turin, Italy\\
$^{56}$ INFN, Sezione di Trieste, Trieste, Italy\\
$^{57}$ Inha University, Incheon, Republic of Korea\\
$^{58}$ Institute for Gravitational and Subatomic Physics (GRASP), Utrecht University/Nikhef, Utrecht, Netherlands\\
$^{59}$ Institute of Experimental Physics, Slovak Academy of Sciences, Ko\v{s}ice, Slovak Republic\\
$^{60}$ Institute of Physics, Homi Bhabha National Institute, Bhubaneswar, India\\
$^{61}$ Institute of Physics of the Czech Academy of Sciences, Prague, Czech Republic\\
$^{62}$ Institute of Space Science (ISS), Bucharest, Romania\\
$^{63}$ Institut f\"{u}r Kernphysik, Johann Wolfgang Goethe-Universit\"{a}t Frankfurt, Frankfurt, Germany\\
$^{64}$ Instituto de Ciencias Nucleares, Universidad Nacional Aut\'{o}noma de M\'{e}xico, Mexico City, Mexico\\
$^{65}$ Instituto de F\'{i}sica, Universidade Federal do Rio Grande do Sul (UFRGS), Porto Alegre, Brazil\\
$^{66}$ Instituto de F\'{\i}sica, Universidad Nacional Aut\'{o}noma de M\'{e}xico, Mexico City, Mexico\\
$^{67}$ iThemba LABS, National Research Foundation, Somerset West, South Africa\\
$^{68}$ Jeonbuk National University, Jeonju, Republic of Korea\\
$^{69}$ Korea Institute of Science and Technology Information, Daejeon, Republic of Korea\\
$^{70}$ Laboratoire de Physique Subatomique et de Cosmologie, Universit\'{e} Grenoble-Alpes, CNRS-IN2P3, Grenoble, France\\
$^{71}$ Lawrence Berkeley National Laboratory, Berkeley, California, United States\\
$^{72}$ Lund University Department of Physics, Division of Particle Physics, Lund, Sweden\\
$^{73}$ Marietta Blau Institute, Vienna, Austria\\
$^{74}$ Nagasaki Institute of Applied Science, Nagasaki, Japan\\
$^{75}$ Nara Women{'}s University (NWU), Nara, Japan\\
$^{76}$ National Centre for Nuclear Research, Warsaw, Poland\\
$^{77}$ National Institute of Science Education and Research, Homi Bhabha National Institute, Jatni, India\\
$^{78}$ National Nuclear Research Center, Baku, Azerbaijan\\
$^{79}$ National Research and Innovation Agency - BRIN, Jakarta, Indonesia\\
$^{80}$ Niels Bohr Institute, University of Copenhagen, Copenhagen, Denmark\\
$^{81}$ Nikhef, National institute for subatomic physics, Amsterdam, Netherlands\\
$^{82}$ Nuclear Physics Group, STFC Daresbury Laboratory, Daresbury, United Kingdom\\
$^{83}$ Nuclear Physics Institute of the Czech Academy of Sciences, Husinec-\v{R}e\v{z}, Czech Republic\\
$^{84}$ Oak Ridge National Laboratory, Oak Ridge, Tennessee, United States\\
$^{85}$ Ohio State University, Columbus, Ohio, United States\\
$^{86}$ Physics department, Faculty of science, University of Zagreb, Zagreb, Croatia\\
$^{87}$ Physics Department, Panjab University, Chandigarh, India\\
$^{88}$ Physics Department, University of Jammu, Jammu, India\\
$^{89}$ Physics Program and International Institute for Sustainability with Knotted Chiral Meta Matter (WPI-SKCM$^{2}$), Hiroshima University, Hiroshima, Japan\\
$^{90}$ Physikalisches Institut, Eberhard-Karls-Universit\"{a}t T\"{u}bingen, T\"{u}bingen, Germany\\
$^{91}$ Physikalisches Institut, Ruprecht-Karls-Universit\"{a}t Heidelberg, Heidelberg, Germany\\
$^{92}$ Physik Department, Technische Universit\"{a}t M\"{u}nchen, Munich, Germany\\
$^{93}$ Politecnico di Bari and Sezione INFN, Bari, Italy\\
$^{94}$ Research Division and ExtreMe Matter Institute EMMI, GSI Helmholtzzentrum f\"ur Schwerionenforschung GmbH, Darmstadt, Germany\\
$^{95}$ Saga University, Saga, Japan\\
$^{96}$ Saha Institute of Nuclear Physics, Homi Bhabha National Institute, Kolkata, India\\
$^{97}$ School of Physics and Astronomy, University of Birmingham, Birmingham, United Kingdom\\
$^{98}$ Secci\'{o}n F\'{\i}sica, Departamento de Ciencias, Pontificia Universidad Cat\'{o}lica del Per\'{u}, Lima, Peru\\
$^{99}$ SUBATECH, IMT Atlantique, Nantes Universit\'{e}, CNRS-IN2P3, Nantes, France\\
$^{100}$ Sungkyunkwan University, Suwon City, Republic of Korea\\
$^{101}$ Suranaree University of Technology, Nakhon Ratchasima, Thailand\\
$^{102}$ Technical University of Ko\v{s}ice, Ko\v{s}ice, Slovak Republic\\
$^{103}$ The Henryk Niewodniczanski Institute of Nuclear Physics, Polish Academy of Sciences, Cracow, Poland\\
$^{104}$ The University of Texas at Austin, Austin, Texas, United States\\
$^{105}$ Universidad Aut\'{o}noma de Sinaloa, Culiac\'{a}n, Mexico\\
$^{106}$ Universidade de S\~{a}o Paulo (USP), S\~{a}o Paulo, Brazil\\
$^{107}$ Universidade Estadual de Campinas (UNICAMP), Campinas, Brazil\\
$^{108}$ Universidade Federal do ABC, Santo Andre, Brazil\\
$^{109}$ Universitatea Nationala de Stiinta si Tehnologie Politehnica Bucuresti, Bucharest, Romania\\
$^{110}$ University of Cape Town, Cape Town, South Africa\\
$^{111}$ University of Derby, Derby, United Kingdom\\
$^{112}$ University of Houston, Houston, Texas, United States\\
$^{113}$ University of Jyv\"{a}skyl\"{a}, Jyv\"{a}skyl\"{a}, Finland\\
$^{114}$ University of Kansas, Lawrence, Kansas, United States\\
$^{115}$ University of Liverpool, Liverpool, United Kingdom\\
$^{116}$ University of Science and Technology of China, Hefei, China\\
$^{117}$ University of Silesia in Katowice, Katowice, Poland\\
$^{118}$ University of South-Eastern Norway, Kongsberg, Norway\\
$^{119}$ University of Tennessee, Knoxville, Tennessee, United States\\
$^{120}$ University of the Witwatersrand, Johannesburg, South Africa\\
$^{121}$ University of Tokyo, Tokyo, Japan\\
$^{122}$ University of Tsukuba, Tsukuba, Japan\\
$^{123}$ Universit\"{a}t M\"{u}nster, Institut f\"{u}r Kernphysik, M\"{u}nster, Germany\\
$^{124}$ Universit\'{e} Clermont Auvergne, CNRS/IN2P3, LPC, Clermont-Ferrand, France\\
$^{125}$ Universit\'{e} de Lyon, CNRS/IN2P3, Institut de Physique des 2 Infinis de Lyon, Lyon, France\\
$^{126}$ Universit\'{e} de Strasbourg, CNRS, IPHC UMR 7178, F-67000 Strasbourg, France, Strasbourg, France\\
$^{127}$ Universit\'{e} Paris-Saclay, Centre d'Etudes de Saclay (CEA), IRFU, D\'{e}partment de Physique Nucl\'{e}aire (DPhN), Saclay, France\\
$^{128}$ Universit\'{e}  Paris-Saclay, CNRS/IN2P3, IJCLab, Orsay, France\\
$^{129}$ Universit\`{a} degli Studi di Foggia, Foggia, Italy\\
$^{130}$ Universit\`{a} del Piemonte Orientale, Vercelli, Italy\\
$^{131}$ Universit\`{a} di Brescia, Brescia, Italy\\
$^{132}$ Variable Energy Cyclotron Centre, Homi Bhabha National Institute, Kolkata, India\\
$^{133}$ Warsaw University of Technology, Warsaw, Poland\\
$^{134}$ Wayne State University, Detroit, Michigan, United States\\
$^{135}$ Yale University, New Haven, Connecticut, United States\\
$^{136}$ Yildiz Technical University, Istanbul, Turkey\\
$^{137}$ Yonsei University, Seoul, Republic of Korea\\
$^{138}$ Affiliated with an institute formerly covered by a cooperation agreement with CERN\\
$^{139}$ Affiliated with an international laboratory covered by a cooperation agreement with CERN.\\

\end{flushleft}

\end{document}